# Advances in Astronomy

## An introduction to a new space debris evolution model - SOLEM

Xiao-wei Wang,[1,2,3] and Jing Liu[1,2]

[1] National Astronomical Observatories, Chinese Academy of Sciences, Beijing, 100101.
[2] Space Debris Observation and Data Application Center, CNSA, Beijing, 100101.
[3] University of Chinese Academy of Sciences, Beijing, 100049.

Correspondence should be addressed to Jing Liu; liujing@bao.ac.cn

**Abstract**

SOLEM is the first space debris long-term evolution model of China. This paper describes the principles, components and workflow of the SOLEM. The effects of different mitigation measures based on SOLEM model are analyzed and presented. The limitation of the model is pointed out and its future improvement work-plan is prospected.

**Introduction**

During the past decades, the number of space objects has been growing rapidly. Until now, the cataloged in-orbit space objects number has reached to about 24,000, about 19000 of which are publicly listed at Space Track [1]. Uncataloged objects number with smaller size has approximately reached to hundreds of millions. These space objects, mostly space debris, pose great threats to operational safety of in-orbit spacecraft. Adopting space debris mitigation measures is an important way to relieve the threats from space debris and prevent the number of resident space objects from growing. However, some studies indicated that the space debris environment would be stable for only 50 years under current mitigation measures, even without new launches in future [2]. This statement has aroused widespread concern over the world. In order to check and quantify the effectiveness of mitigation measures on controlling the growth of space debris in future, many space debris evolution models are established and compared to study the long-term stability of the future space environment.

At present, the well-known space debris evolution models mainly include the LEGEND model from National Aeronautics and Space Administration (NASA) [3], DAMAGE model from United Kingdom Space Agency (UKSA) [4], MEDEE model from Centre National d'Etudes Spatiales (CNES) [5], DELTA model from European Space Agency (ESA) [6], LUCA model from Technische Universität Braunschweig [7], NEODEEM model from Kyushu University and The Japan Aerospace Exploration Agency (JAXA)[8], etc. Some of these models have been used to study the stability of the future space environment in the joint research organized by Inter-Agency Space Debris Coordination Committee (IADC) [9,10]. Besides, further work on the uncertainties affecting the long-term evolution of space debris is encouraged in international community to better asses the uncertainty induced by the modelling assumptions [11]. Therefore, more space debris evolution models are welcomed to participate in such research activities, which may provide the technical supports for making new space debris mitigation guidelines as well as other related policies for space traffic management to guarantee the long-term sustainability of outer space activities.

SOLEM (Space Objects Long-term Evolution Model) is a Low Earth Orbit (LEO) space debris long-term evolution model established by China. It has participated in the joint researches of IADC as a representative of China National Space Administration (CNSA). SOLEM is capable of predicting the number evolution trends of space debris, estimating the rate of collision events of space objects during the evolution in future, and analyzing the effects of different mitigation and remediation measures or other potential uncertainties on the long-term evolution of space debris. The reliability of SOLEM has been validated during the joint research of IADC.

This paper introduces the components, algorithms and workflow of SOLEM. After that, the effects of different mitigation measures based on SOLEM model are analyzed.

**The SOLEM Model**

The space debris evolution model is expected to predict the evolution of space debris population and possible collision rates for a long period in future, usually for decades even centuries. It can be used to study the evolution processes with various assumptions. The future evolution of space debris is affected by natural factors such as various perturbations, atmosphere evolutions, periodic solar activities, accidental explosions, and even the surface degradations. In fact, it could also be affected by human space activities such as launches, collision avoidance manoeuvres, mitigation and remediation measures. In space debris evolution model, usually the most important source and sink mechanisms are considered. Generally, a space debris evolution model is composed of orbital propagation model, collision probability estimation model, fragment generation model, future launch model, post mission disposal model and active debris removal model (if the active debris removal measures are considered). These components will significantly affect the model evolution results if some key parameters are changed. The composition of space debris evolution model is illustrated in Figure 1.

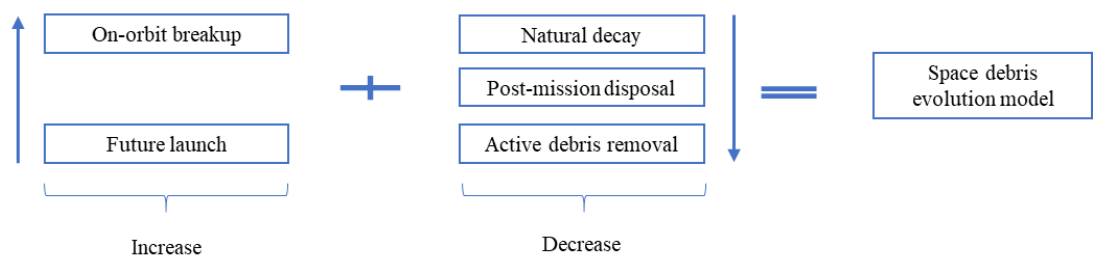

Figure 1: The general components of space debris evolution models. The left components are the main source mechanisms, and right components are the main sink mechanisms.

# Orbital propagation

Orbital propagation is to project the current orbits of space objects to the future. It is the core component of space debris evolution model. Through orbital propagation, the space debris evolution model is able to obtain the space objects orbital distribution at any moment in future. There are three basic orbital prediction algorithms: numerical method, analytical method and semi-analytical method. Numerical method has the highest precision, but takes the most time in orbit propagation. Due to the long evolution time of space debris, usually from decades to hundreds of years, moreover, the high-precision position has no practical significance in long-term evolution, it is more appropriate to use analytical method or semi-analytical method.

SOLEM model adopts a simplified semi-analytical orbital propagator, in which the integration is done on the perturbation functions with the short-periodic terms removed. Essentially, it is performed on the averaged orbital dynamic system. At present, SOLEM covers only LEO region, including objects residing in LEO with near-circular orbits and those crossing LEO with high eccentricity orbits. For near-circular orbits, the main perturbations considered includes the Earth's non-spherical gravity perturbation $J_2$, $J_3$, $J_4$, $J_{2,2}$, and atmospheric drag. For high eccentricity orbits, besides the Earth's non-spherical gravity and atmospheric drag, the perturbations due to solar radiation pressure and gravity of the Sun and Moon are also considered. The atmosphere density model used for drag calculation is the NRLMSIS00 model. The values of solar radiation flux at 10.7 cm and the geomagnetic index can be read from a configuration file which can be replaced according to assumptions.

In order to verify this orbital propagator of SOLEM, we conducted an experiment on the evolution of a small population. It is to compare the SOLEM propagation results with historical data for the number evolution of a small population in a statistical view. We used all the 1021 cataloged LEO-crossing objects on 1980.01.01 to do the experiment. It includes 38 objects with high eccentricity orbits ($e \geq 0.2$) and 983 objects with near circular orbits ($e < 0.2$). The area-to-mass ratio of these objects are calculated according to the **UNW** type of perturbed motion equation together with the method of least squares, using the orbital data for months previously. For SOLEM propagation, we used historical solar activities recorded in CelesTrak website [12] and considering no collision avoidance and station keeping manoeuvres. The real decay information of the 1021 objects are drawn from SSR on the space-track website [13]. The propagation result of SOLEM orbital propagator and the real data of historical evolution of the 1021 objects are compared in Figure 2, which shows a high consistency with a relative error of about 2%.

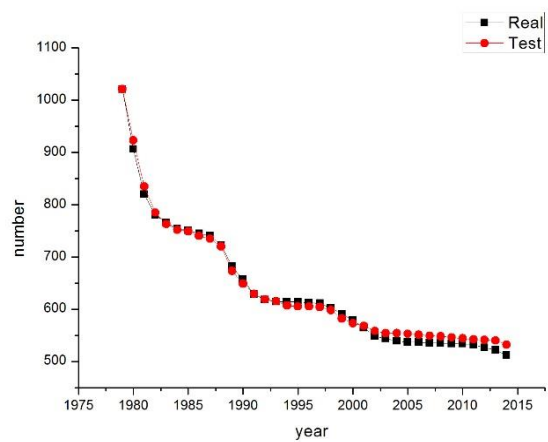

Figure 2: The statistical results comparison of SOLEM propagation (noted as Test) and historical evolution (noted as Real).

The semi-analytical method has a limit precision in orbit propagation. However, comparing with the evolution of a single orbit, the space debris long-term evolution model cares more about the number evolution of the whole population in statistics. Considering the experiment above, we think the SOLEM orbital propagator is applicable to space debris long-term evolution model.

**Fragment generation model**

In-orbit breakup is one important source of space debris growth. Therefore, the accuracy of fragment generation model simulating the breakup events has an important impact on the simulation results of space debris evolution model. The fragment generation model is to simulate the space debris collisions or explosions and give the instantaneous information of generated fragments which is necessary for the subsequent evolution prediction. The information includes the fragments number, each fragment's mass, size, velocity, etc.

In SOLEM, we adopt NASA's standard breakup model to simulate the generation of fragments produced by in-orbit breakups. NASA's standard breakup model is the most popular fragment generation model at present. The implementation is following the process presented in paper [14,15].

**Collision probability estimation**

When considering the fragmentation due to in-orbit collisions, there is a key component in the space debris evolution model, that is the collision probability estimation algorithm. In SOLEM, we adopt an Improved-CUBE (I-CUBE) model to do the calculation of collision probabilities [16]. It is based on the CUBE method proposed by NASA [17,18].

In CUBE model, the evolution system is uniformly sampled in time. At each sampling moment, the space around the Earth is discretized in small cubes in geocentric cartesian coordinates. By obtaining updated orbital elements, the location of each space objects is calculated. CUBE model assumes that the collision probability only exists between objects residing in the same cube. And the collision probability is calculated by

$$P_{ij} = s_i s_j A_c V_{imp} dU dt \tag{1}$$

where $s_i$ and $s_j$ are the spatial densities of objects $i$ and $j$ in the cube, $A_c$ is the collision cross-section, $V_{imp}$ is collision speed, $dU$ is the volume of the cube, and $dt$ is the time interval between two sampling moments.

Actually, $P_{ij}$ calculated by Equation 1 is the mean number of collisions between objects $i$ and $j$ in the volume $dU$ during the propagation time interval $dt$. The time interval $dt$ is given as 5 days, i.e. $4.32 \times 10^5$ seconds. As it does not approach to 0, for some objects with collision cross-section $A_c$ large enough, the value of $P_{ij}$ will reach to greater than 1. That is not reasonable. To avoid this, in I-CUBE model, we used Equation 2 to express the collision probability with the consideration that the collision process follows a Poisson distribution.

$$\begin{cases} P_{ij} = 1 - \exp(-c) \\ c = s_i s_j A_c V_{imp} dU dt \end{cases} \tag{2}$$

where $P_{ij}$ represents the collision probability, and $c$ is the mean number of collisions between objects $i$ and $j$ in the volume $dU$ during the propagation time interval $dt$.

According to Heiner Klinkrad [19], the approximation $P_{ij} \approx c$ yields results with less than 10% error for $c \leq 0.2$. That means for $c > 0.2$, the approximation will bring error bigger than 10%. For most space objects, the approximation is well suited. But for those with collision cross-sections large enough (dozens or even hundreds of square meters), the collision probability may be greatly over estimated if still using the approximation.

Besides, CUBE model assumes that only the objects residing in the same cube are considered for collisions. For space debris evolution, the divided cube size is given as 10 km. However, it has been queried by CNES for the effects on evolution results from the divided cube size [20,21]. In I-CUBE model, we assume that collision probability exists in all close approaches with a distance from the target satisfying the threshold. The distance threshold is the diagonal of the divided cube. Thus, the value of $dU$ in Equation 1 is no longer the volume of cube, but the volume of a sphere with radius equal to the distance threshold, i.e.

$$dU = \frac{4\pi}{3}\left(\sqrt{3}h\right)^3 = 4\pi \bullet \sqrt{3}h^3 \qquad (3)$$

Where $h$ is the divided cube size. As $dU$ relates to the spatial densities, $s_i$ and $s_j$ are now the spatial densities of objects $i$ and $j$ in the volume of the sphere. The two-dimensional representation is illustrated in Figure 3.

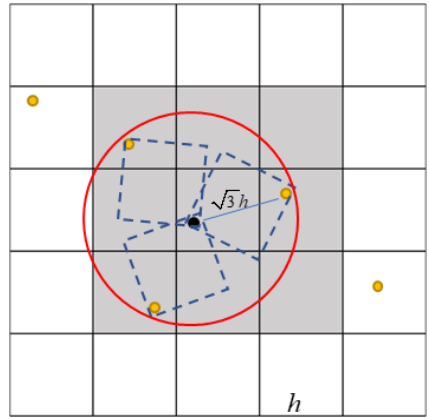

Figure 3: Two-dimensional representation for considering possible collisions between debris residing in neighbouring cubes.

In this approach, the divided cube size will never influence the evolution result of space debris evolution models. The comparison results using CUBE and I-CUBE model running by SOLEM are presented in Figure 4. The divided cube size $h$ varies from 5 km to 50 km. Except for the divided cube size, all the other configurations are the same. Every curve is the average result of 50 Monte Carlo runs.

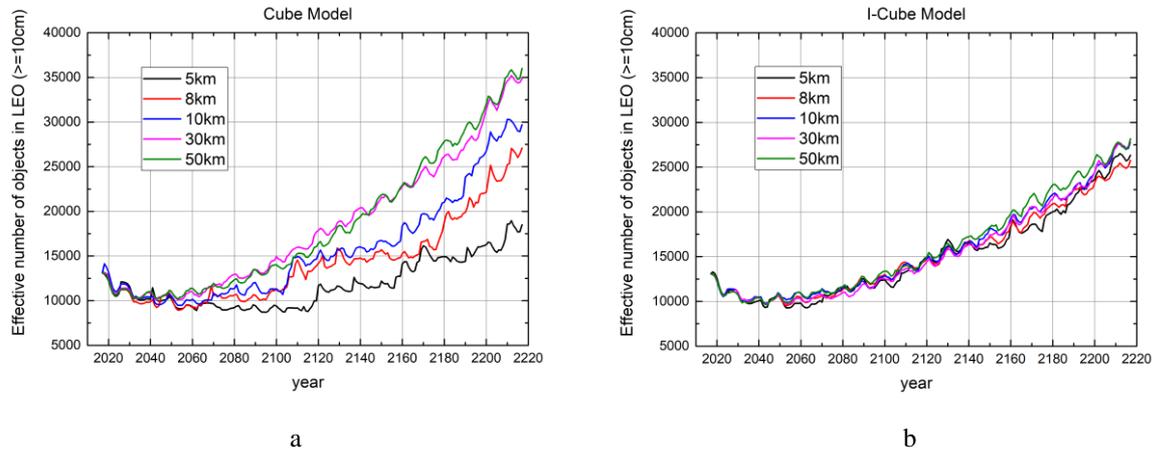

a                                                                  b

Figure 4: Comparison of simulation results with different cube size. (a) Using CUBE model. (b) Using I-CUBE model

**Future launch activities**

The launch of spacecraft in future is another important source of space debris increase. However, it is highly related with technical development and space policies which can't be predicted. Therefore, the future launch model usually takes the current launch level as a reference. The data of a launch model includes all the characteristics of launched objects, such as the launched number, each object's type, mass, area or/and size, target orbit, launch time, etc.

In SOLEM model, we adopt the launch traffic during the last 8 years, from September 1, 2009 to August 31, 2017, as future launch model. It will be repeated during the overall simulation time. The traffic data is collected mainly from websites of Space Launch Report [22], Space Track [23] and Union of Concerned Scientists [24]. It is prepared previously as a configuration file containing the information of launched numbers, types (including satellites, rocket bodies and mission related objects), each object's mass, area (or/and size), target orbit, launch date, etc.

**Post-mission disposal**

Post-Mission Disposal (PMD) is an important mitigation measure to stop space debris population from growing. In SOLEM model, PMD measures are implemented on non-functional satellites and rockets launched during the evolution time. For newly launched satellites, the mission life is uniformly set as 8 years by default. It can also be set as other values by user. For rockets, the mission life will end at once when the carried satellites are sent into the target orbits. When the mission life of a satellite or rocket ends, the natural orbital lifetime will be estimated. If the natural orbital lifetime exceeds 25 years, the satellite or upper stage of the rocket will be deorbited to a disposed orbit that will naturally decay within 25 years, complying with the 25-year rule. The PMD success rate in SOLEM can be set freely by users. Currently this value is estimated lower than 20% for region above 600 km. The procedure of PMD is showed in Figure 5.

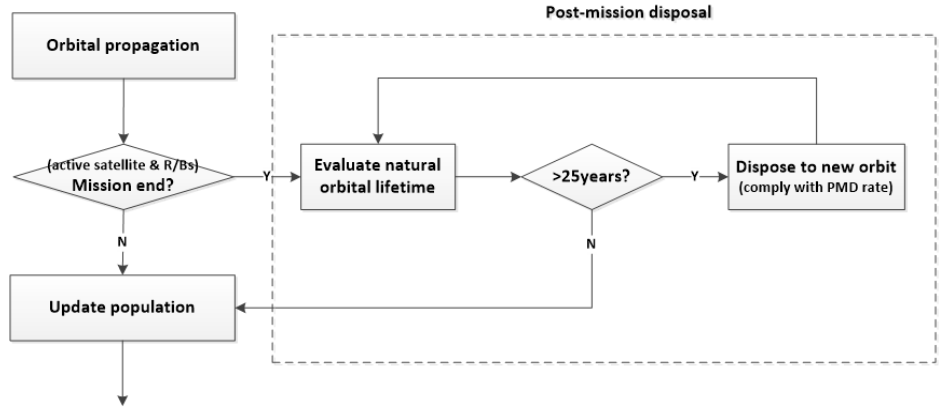

Figure 5: The procedure of PMD. For mission ended satellites or rockets (R/Bs), if the evaluated natural orbital lifetime excesses 25 years, it will be disposed to a new orbit complying the 25-year rule.

**Active debris removal**

To better limit the growth of LEO space debris populations, measures of Active Debris Removal (ADR) are suggested. Although the ADR has not become practical due to the technical difficulties and high costs, its effects on space debris evolution has been proved through computer simulations. Considering the developing technology, ADR will be another important measure in stopping the growth of the space debris population in future. As suggested, ADR measure is to remove existing large and massive objects from regions where high collision activities are expected [25]. The selection criteria that should be used in choosing which objects to remove has also been researched, and the criterion based on the mass and collision probability of each object has been proposed [26,27,28]. By annually removing several targets, the space environment can be stabilized according to computer simulations.

In SOLEM model, the selection criterion is implemented as follows:

$$R_i = \sum p_{ij} \times m_i \qquad (8)$$

where $m_i$ is the mass of object $i$, $\sum p_{ij}$ is the cumulated collision probabilities between object $i$ and object $j$ where $j \neq i$ during the last year. Their product $R_i$ is the selection index for ADR. The larger the value of $R_i$, the more dangerous the object $i$.

At the beginning of each projection year, all objects in orbit are sorted in descending order by the value of $R_i$. A pre-defined number of space debris with the largest $R_i$s will be immediately removed from orbits. Only the operating satellites and objects with high eccentricity orbits are excluded. The beginning year of implementing ADR measures is set by users. In SOLEM, it is set as 2030 by default.

## The initial population

Space objects initial population is the baseline of space debris evolution model. It is the description of current space environment. For SOLEM, the population data on 2017.09.01 is used as initial population. Just like the future launch model, the information of Space objects is obtained from Space Track, Space Launch Report and Union of Concerned Scientists. The orbital distribution and the Area-to-Mass Ratio (A/M) VS. size distribution are shown in Figures 6 and 7.

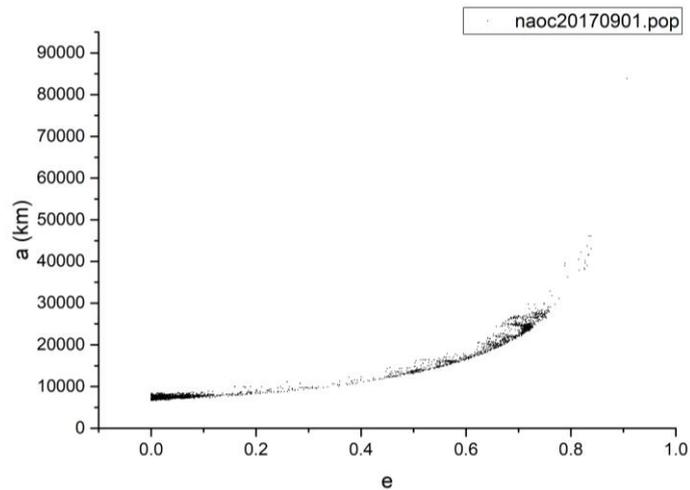

Figure 6: The semi-major axis VS. eccentricity distribution of population data of 2017.09.01.

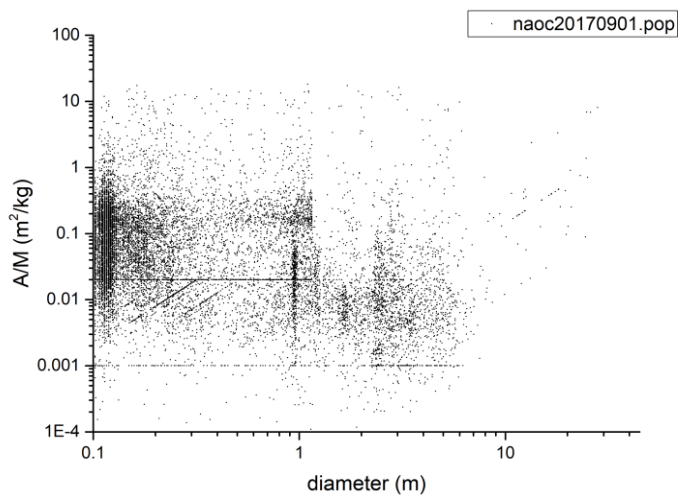

Figure 7: The A/M VS. size distribution of population data of 2017.09.01.

## The workflow of SOLEM model

The workflow of SOLEM model is simply represented by Figure 8. As presented, before projection, initialization will be done first by setting key parameters which are based on simulated assumptions and taking prepared initial population data as input. All space objects contained in the initial population are propagated after initialization. As time evolves, the newly launched objects from future launch model, will also be propagated. If the newly launched active satellite or rocket ends its mission, the PMD measure will be done. All space objects with size over 10 cm are included for collision consideration. Once a collision happens, the breakup model will be used to generate new fragments. And the population for next propagation step will be updated.

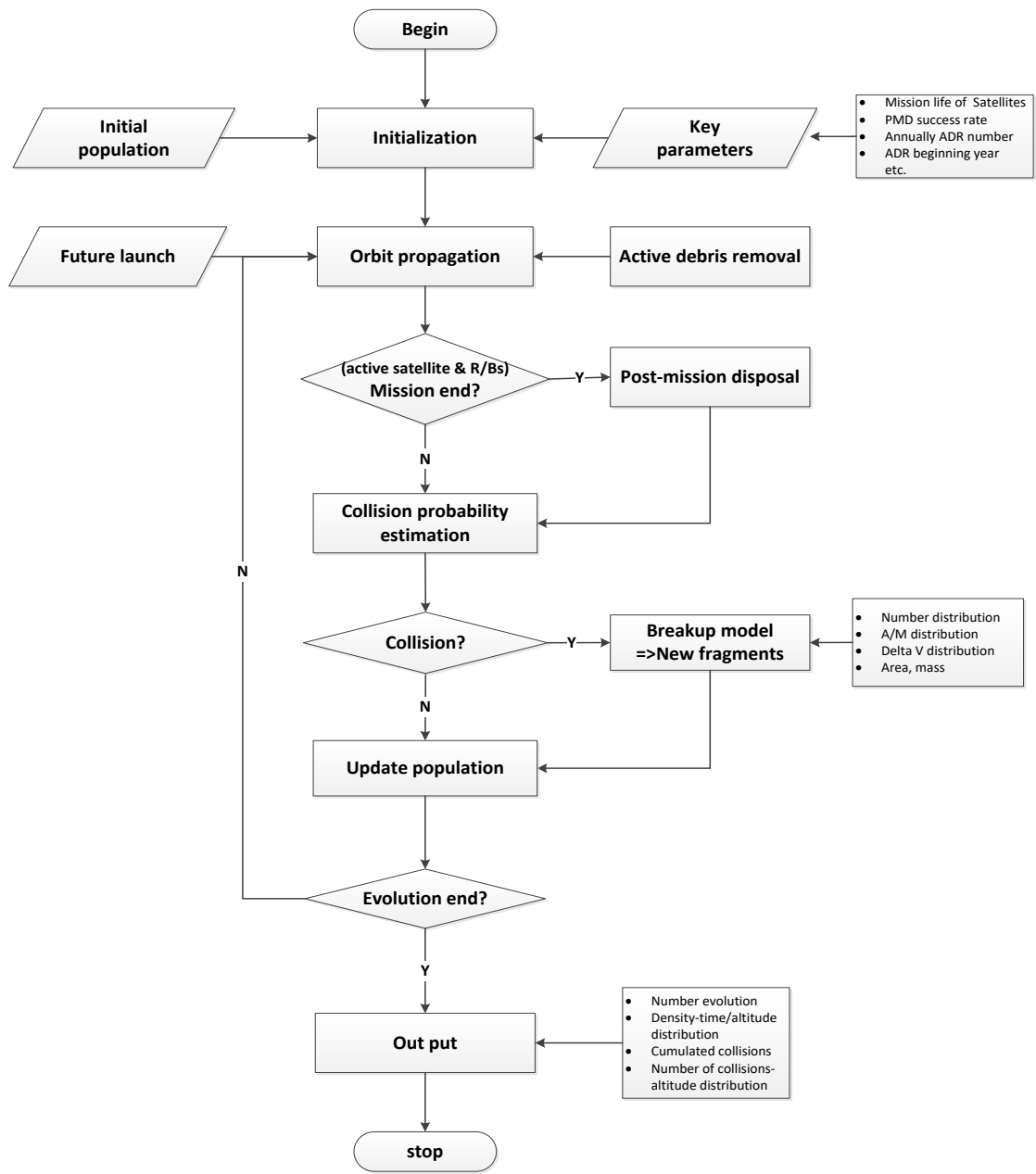

Figure 8: The workflow of SOLEM model.

**Model Application**

As key parameters of each module are flexible to users, SOLEM model is able to simulate the evolution of space debris under various assumptions with high flexibility. Since 2015, SOLEM, as a representative of CNSA, has participated in a joint research of IADC. With uniform input data and assumptions, SOLEM has achieved results consistent with other space debris evolution models (IADC internal reports). In this paper, the effects of different mitigation measures on space debris evolution are analyzed with the SOLEM model.

## Input Data

The initial input data and relevant assumptions are shown in Table 1. Three scenarios are performed with PMD rate setting as 30%, 60% and 90%, and the other input data and assumptions are all the same. For each scenario, 50 Monte Carlo simulation runs are performed to obtain the averages.

Table 1. Assumptions of scenarios simulated by SOLEM model.

| Items | Descriptions |
|---|---|
| Initial population data | 2017.09.01, 10 cm and larger objects, crossing LEO |
| Future launch model | The launch traffic from 2009.09.01 to 2017.08.31, repeated during overall simulation time |
| Satellite mission life | 8 years |
| Collision avoidance maneuvers | no |
| Station keeping | no |
| Passivation rate | 100% |
| PMD rate | 30%, 60%, 90% |
| ADR | no |
| Collision types | Collision energy E>40 J/g, catastrophic collisions; else, non-catastrophic collisions |
| Evolution time | 200 years |
| MC runs | 50 MC |

The solar activity used in SOLEM for future evolution is shown in Figure 9. It is generated according to the monthly fit formula offered by CelesTrak website [12]. The geomagnetic index is set as a constant median value of Ap=9.

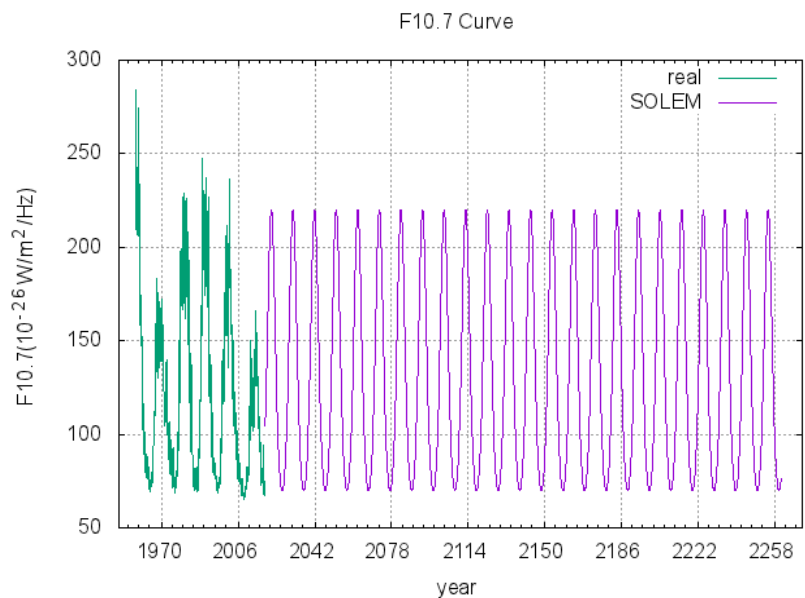

Figure 9: The solar activity recorded in history (green line, noted as real) and the solar activity model adopted in SOLEM (purple line).

## Simulation Results

In the evolution results, space objects are classified into three types: (1) Intacts, includes all satellites, R/Bs, and mission-related objects; (2) Old fragments, means all the DEB already existed in the initial population; (3) New fragments, means all the DEB generated during the evolution time. Separating new fragments from old fragments can help us have a clear view of the increasing process of space debris population.

The space debris evolution results of the scenario setting PMD rate as 30% is presented in Figure 10. It is the average result of 50 Mont-Carlo runs by SOLEM. As Figure 10 shows, the total number of objects in LEO shows a decrease in the first two decades and then turns into increase throughout the evolution time and finally reaches to more than 115% of the initial population. This scenario predicts 34 catastrophic collisions and 25 non-catastrophic collisions in average in future 200 years.

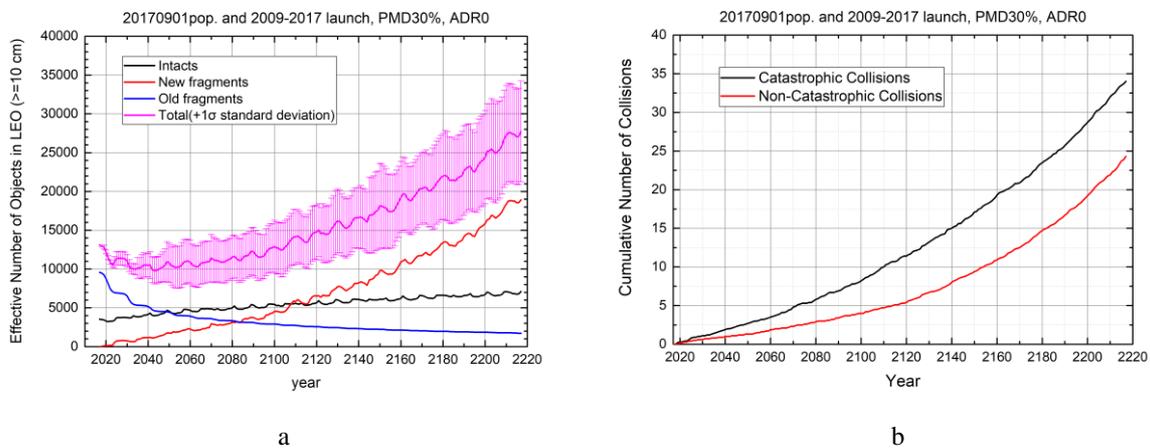

Figure 10: The evolution results of scenario 1, with PMD rate of 30%. (a) The population evolution. The line of Total is plot with the error bar of $1\sigma$ standard deviation. (b) The cumulative number of collisions.

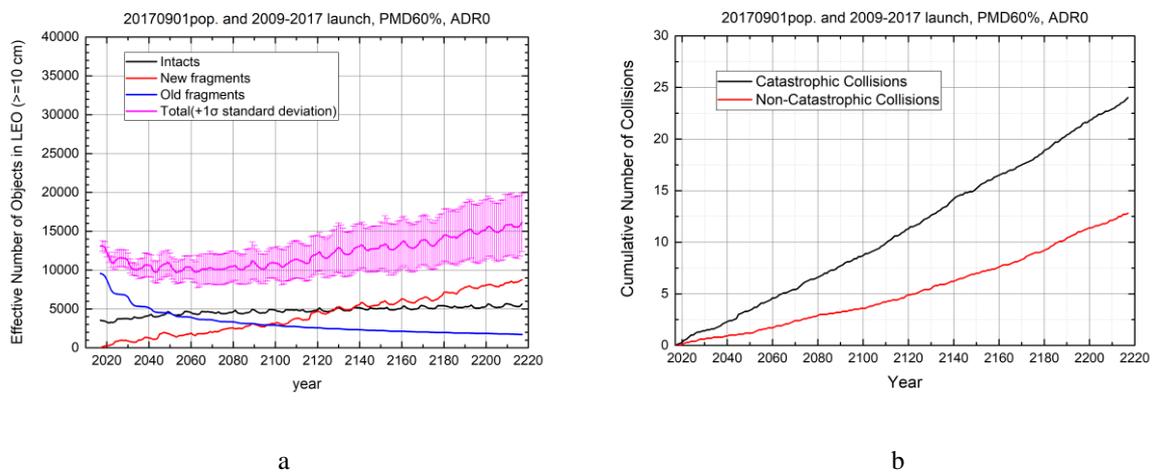

Figure 11: The evolution results of scenario 2, with PMD rate of 60%. (a) The population evolution. The line of Total is plot with the error bar of $1\sigma$ standard deviation. (b) The cumulative number of collisions.

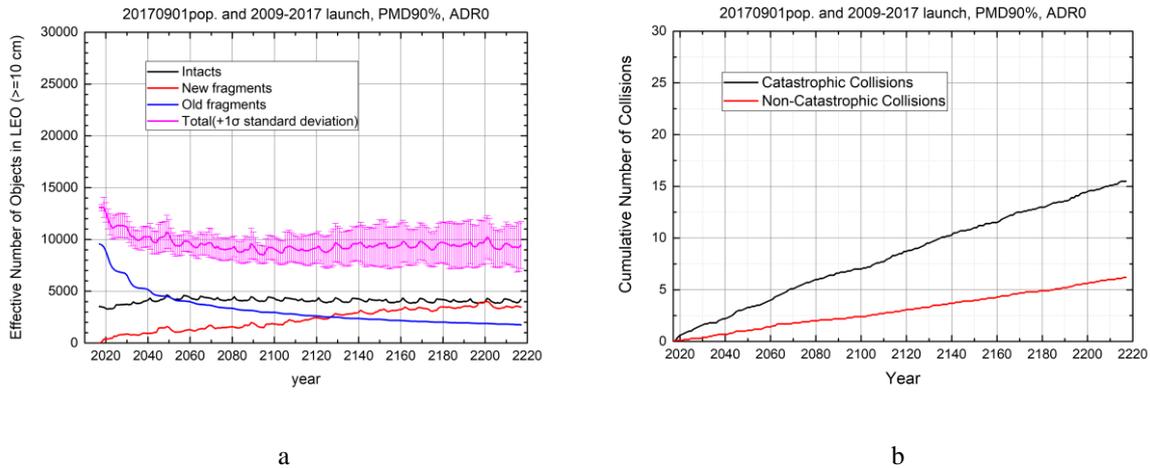

Figure 12: The evolution results of scenario 3, with PMD rate of 90%. (a) The population evolution. The line of Total is plot with the error bar of 1σ standard deviation. (b) The cumulative number of collisions.

Figure 11 shows the evolution results of the scenario setting PMD rate as 60%. The reinforcement of such mitigation measure making the final effective number of LEO objects in future 200 years decrease greatly comparing with the baseline scenario. The final total effective number of LEO objects is only 23% more than the initial population. And the cumulative number of collisions also decrease greatly in both collision types.

In Figure 12, the evolution result shows, with PMD rate of 90%, there is a clear decrease by approximately 30% in total effective number of space objects crossing LEO orbits for next 50 years, and then the population remains at a long-term stable level. The decrease in the first 50 years is mainly due to the natural decay of old fragments. The number of new fragments generated by breakup events increases in nearly the whole evolution time with a low rate, and finally seems to stop increasing at the end of evolution. The cumulative number of catastrophic collisions is decreased down to 15, and for non-catastrophic collisions the number is only 7. Generally, this scenario predicts a space debris environment becoming better with PMD rate as high as 90%.

Simulation results of the three scenarios are quantified in Table 2. It can be seen that, with PMD rate increases, the space debris population after 200 years will greatly decrease, as well as the average catastrophic collision rates. High PMD rates will make the current space environment become better and safer.

Table 2. Quantification of evolution results of the three scenarios simulated by SOLEM model.

| simulations | PMD rate | % of Variation wrt Initial population after 200 years | Average catastrophic collision rate in future |
|---|---|---|---|
| Scenario 1 | 30% | +115% | 1/6 years |
| Scenario 2 | 60% | +23% | 1/8 years |
| Scenario 3 | 90% | -30% | 1/13 years |

Taking the IADC comparison study about "Stability of the Future Leo Environment" [9,10] as a reference, the evolution results shown above look rather optimistic. The IADC comparison study predicted about +30% changes in population after 200 years and one catastrophic collision every 5 to 9 years with PMD rate of 90%. And we predict -30% change in population and one catastrophic collision every 13 years with the same PMD compliance level.

That might be mainly due to the differences in solar activity model and the input initial population used for simulation. The solar activity used in this paper (Figure 9) is in a higher level than those used in references [9,10] which is shown in Figure 13. This will make more objects decay during the evolution. Besides, the initial population we used in this paper is obtained from the public data on 2017.09.01, which is about 13000 space objects. While the initial population used in references [9,10] is the reference population of MASTER2009 on 2009.05.01, which is about 17000 space objects. The difference in initial population is as high as about 24%. Additionally, the area-to-mass ratio distribution of the initial population in this paper (Figure 7) is also different from reference [9,10] which is shown in Figure 14. From the area-to-mass ratio distribution of the initial population, it can be seen that the initial population we used do not exclude those objects with high area-to-mass ratio.

The differences in solar activity projection and initial population including both the number and area-to-mass ratio finally leads to a very different evolution result.

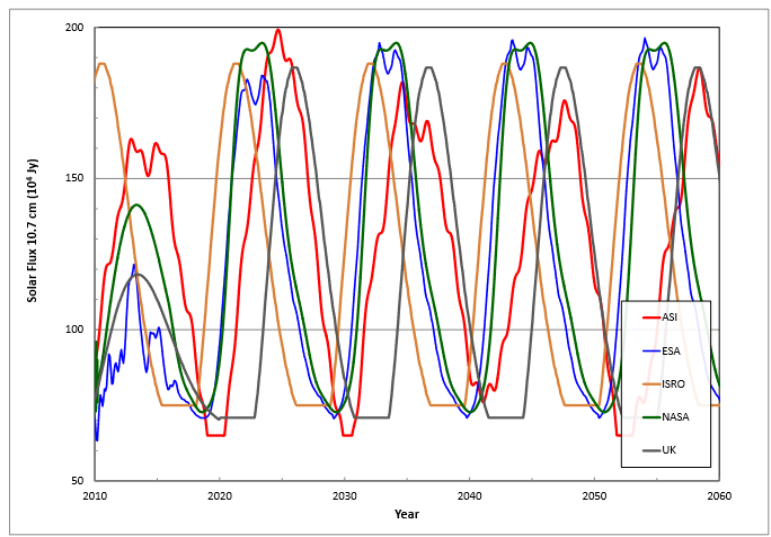

Figure 13: Solar flux projections used in IADC comparison study.

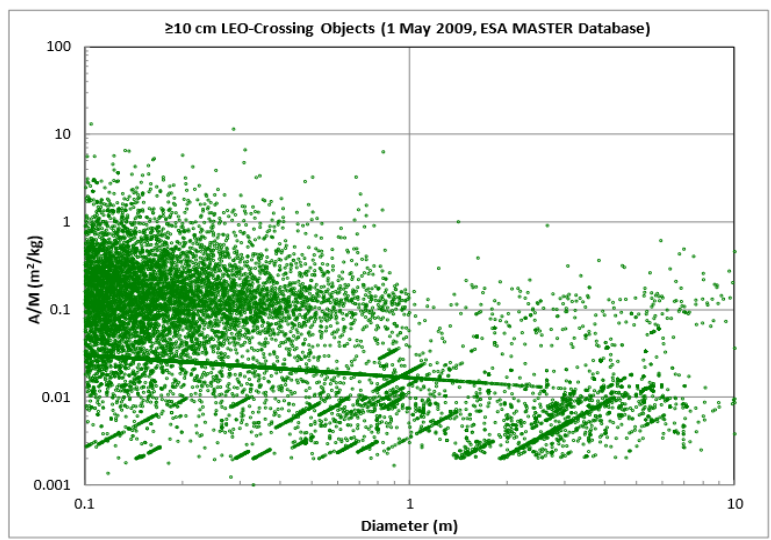

Figure 14: Area-to-Mass ratio distributions of the initial population used in IADC comparison study.

**Summary and future work**

This paper mainly introduced the composition, sub-model algorithm, and workflow of SOLEM, the space debris long-term evolution model of China. The reliability of SOLEM has been validated during the joint research of IADC. After that, the application work of SOLEM model on analyzing the effects of different mitigation measures on the evolution of space environment are presented. Result shows with higher PMD rate, the current space environment will become better and safer.

SOLEM is a LEO space debris evolution model with high flexibility. It is capable of simulating the space environment evolution with various assumptions. Therefore, it can be used to simulate and analyze the uncertainties affecting the space debris evolution, such as the future launches, solar activities, manual collision avoidance measures, mitigation and remediation measures, and so on. Through simulation and analyzation, SOLEM can help us to deeply understand the evolution process of space environment and provides technical supports for making space policies and laws to guarantee the sustainability of space activities in future.

At present, the orbital range covered by SOLEM is limited to LEO region from 200 km to 2000 km. In the next step, the orbital range covered by SOLEM will be expanded from LEO region to GEO (Geostationary Earth Orbit) region. Besides, the post-mission disposal model will be optimized, including the disposed orbit selection process and the computation time.

**Conflicts of Interest**

No.

**Funding Statement**

The work presented in this paper is supported by Natural Science Foundation of China (grant number 11503044) and Space Debris Research Project (grant numbers KJSP2016010101, KJSP2016020201, KJSP2016020301, KJSP2016020101).

**Acknowledgments**

Thanks to Cui Shuang-xing, Cheng Hao-wen, Zhang Yao, Yang Zhi-tao, Shen Dan, Wu Xiang-bin and other colleagues for the discussion and help during the establishment of SOLEM model.